\documentclass[showpacs,showkeys,twoside,eqsecnum]{revtex4}

\usepackage{amsfonts,amsmath,amssymb,bm,graphicx}

\begin{document}


\title{Formation of Cooper pairs in quantum oscillations of electrons in plasma}

\author{Maxim Dvornikov}
\affiliation{Departamento de F\'{i}sica y Centro de Estudios
Subat\'{o}micos, Universidad T\'{e}cnica Federico Santa Mar\'{i}a,
Casilla 110-V, Valpara\'{i}so, Chile}
\email{maxim.dvornikov@usm.cl}
\affiliation{IZMIRAN, 142190, Troitsk, Moscow region, Russia}

\date{\today}

\begin{abstract}
We study low energy quantum oscillations of electron gas in
plasma. It is shown that two electrons participating in these
oscillations acquire additional negative energy when they interact
by means of a virtual plasmon. The additional energy leads to the
formation a Cooper pair and possible existence of the
superconducting phase in the system. We suggest that this
mechanism supports slowly damping oscillations of electrons
without any energy supply. Basing on our model we put forward the
hypothesis the superconductivity can occur in a low energy ball
lightning.
\end{abstract}

\pacs{52.35.-g, 74.20.Mn, 92.60.Pw}

\keywords{plasmon, electron gas oscillations, Cooper pairs,
superconductivity, ball lightning}

\maketitle

\section{Introduction}\label{INTR}

In the majority of cases plasma contains a great deal of
electrically charged particles. Free electrons in metals can be
treated as a degenerate plasma. If the motion of charged particles
in plasma is organized, one can expect the existence of
macroscopic quantum effects. For instance, collective oscillations
of the crystal lattice ions can be exited in metals. These
oscillations are interpreted as quasi-particles or phonons.
Phonons are known to be bosons. The exchange of a phonon causes
the effective attraction between two free electrons in plasma of
metals. This phenomenon is known as the formation of a Cooper
pair~\cite{Coo56} and is important in the explanation of the
superconductivity of metals. Cooper pairs can exist during a
macroscopic time resulting in an undamped electric current.
Unfortunately the formation of Cooper pairs is possible only at
very low temperatures. Nowadays there are a great deal of attempts
to obtain a superconducting material at high
temperatures~\cite{Gin00}.

Without an external field, e.g. a magnetic field, the motion of
electrons in plasma is stochastic. It results in rather short
times of plasma recombination. The laboratory plasma at
atmospheric pressure without an energy supply recombines during
$\sim 10^{-3}\thinspace\text{s}$~\cite{Smi93}. That is why free
oscillations of electrons in plasma, i.e. without any external
source of energy, will attenuate rather fast. Therefore to form a
Cooper pair in plasma one should create a highly structured
electrons motion to overcome the thermal fluctuations of
background charged particles. Nevertheless it is known that a
plasmon superconductivity can appear~\cite{Pas92}.

In Ref.~\cite{Dvo} we studied quantum oscillations of electron gas
in plasma on the basis of the solutions to the non-linear
Schr\"odinger equation. The spherically and axially symmetrical
solutions to the Schr\"odinger equation were found in our works.
We revealed that the densities of both electrons and positively
charged ions have a tendency to increase in the geometrical center
of the system. The found solutions belong to two types, low and
high energy ones, depending on the branch in the dispersion
relation (see also Sec.~\ref{MODEL}). We put forward a hypothesis
that a microdose nuclear fusion reactions can happen inside a high
energy oscillating electron gas. This kind of reaction can provide
the energy supply to feed the system. In our works we suggested
that the obtained solutions can serve as a model of a high energy
ball lightning. Note that the suggestion that nuclear reactions
can take place inside a ball lightning was also put forward in
Ref.~\cite{nuclfus}.

In the present work we study a low energy solution (see
Sec.~\ref{MODEL}) which is also presented in our model~\cite{Dvo}.
As we mentioned above a non-structured plasma usually recombines
during several milliseconds. It is the main difficulty in
constructing of long-lived plasma structures. In a high energy
solution internal nuclear fusion reactions could compensate the
energy losses. We suggest that the mechanism, which prevents the
decay of low energy oscillations of the electron gas, is based on
the possible existence of the plasma superconductivity. In frames
of our model we show that the interaction of two electrons,
described by spherical waves, can be mediated by a virtual
plasmon. For plasmon frequencies corresponding to the microwave
range, this interaction results in the additional negative energy.
Then it is found that for a low energy solution with the specific
characteristic this negative energy results in the effective
attraction. Therefore a Cooper pair of two electrons can be formed
because the interacting electrons should have oppositely directed
spins. Using this result it is possible to state that the friction
occurring when electrons propagate through the background matter
can be significantly reduced due to the superconductivity
phenomenon. We also discuss the possible applications of our
results for the description of the a low energy ball lightning. It
should be noticed that the appearance of the superconductivity
inside of a ball lightning was considered in
Ref.~\cite{supercondold}.

This paper is organized as follows. In Sec.~\ref{MODEL} we briefly
outline our model of quantum oscillations of the electron gas
which was elaborated in details in Ref.~\cite{Dvo}. Then in
Sec.~\ref{LE} we study the interaction of two electrons when they
exchange a virtual plasmon and examine the conditions of the
effective attraction appearance. In Sec.~\ref{CONCL} we summarize
and discuss the obtained results.

\section{Brief description of the model}\label{MODEL}

In Ref.~\cite{Dvo} we discussed the motion of an electron with the
mass $m$ and the electric charge $e$ interacting with both
background electrons and positively charged ions. The evolution of
this system is described by the non-linear Schr\"odinger equation,
\begin{equation}\label{Schrod0}
  \mathrm{i}\hbar\frac{\partial \psi}{\partial t} = H_0 \psi,
  \quad
  H_0 = -\frac{\hbar^2}{2m}\nabla^2+U(|\psi|^2),
\end{equation}
where
\begin{equation}\label{Uinterac}
  U(|\psi|^2)=e^2\int\mathrm{d}^3\mathbf{r}'
  \frac{1}{|\mathbf{r}-\mathbf{r}'|}
  \{|\psi(\mathbf{r}',t)|^2-n_i(\mathbf{r}',t)\},
\end{equation}
is the term which is responsible for the interaction of an
electron with background matter which includes electrons and
positively charged ions with the number density
$n_i(\mathbf{r},t)$.  In Eq.~\eqref{Schrod0} the wave function is
normalized on the number density of electrons,
$|\psi(\mathbf{r},t)|^2=n_e(\mathbf{r},t)$.

Note that in Eq.~\eqref{Schrod0} the $c$-number wave function
$\psi = \psi(\mathbf{r},t)$ depends on the coordinates of a single
electron. Generally in a many-body system this kind of
approximation is valid for an operator wave function, i.e. when it
is expressed in terms of creation or annihilation operators. In
Ref.~\cite{KuzMak99} it was shown that it is possible to rewrite
an exact operator Schr\"odinger equation using $c$-number wave
functions for a single particle. Of course, in this case a set of
additional terms appears in a Hamiltonian. Usually we neglect
these terms (see Ref.~\cite{Dvo}).

The solution to Eqs.~\eqref{Schrod0} and~\eqref{Uinterac} can be
presented in the following way:
\begin{equation}\label{sol0}
  \psi(\mathbf{r},t) = \psi_0+\delta\psi(\mathbf{r},t),
  \quad
  |\psi_0|^2 = n_0,
\end{equation}
where $n_0$ is the density on the rim of system, i.e. at rather
remote distance from the center. The perturbative wave function
$\delta\psi(\mathbf{r},t)$ for the spherically symmetrical
oscillations has the form (see Ref.~\cite{Dvo}),
\begin{equation}\label{pertsol}
  \delta\psi(\mathbf{r},t) = e^{-\mathrm{i}\omega t}
  \delta\psi_k(\mathbf{r}),
  \quad
  \delta\psi_k(\mathbf{r}) = A_k \frac{\sin kr}{r},
\end{equation}
where $A_k$ is the normalization coefficient.

\begin{figure}
  \centering
  \includegraphics[scale=.5]{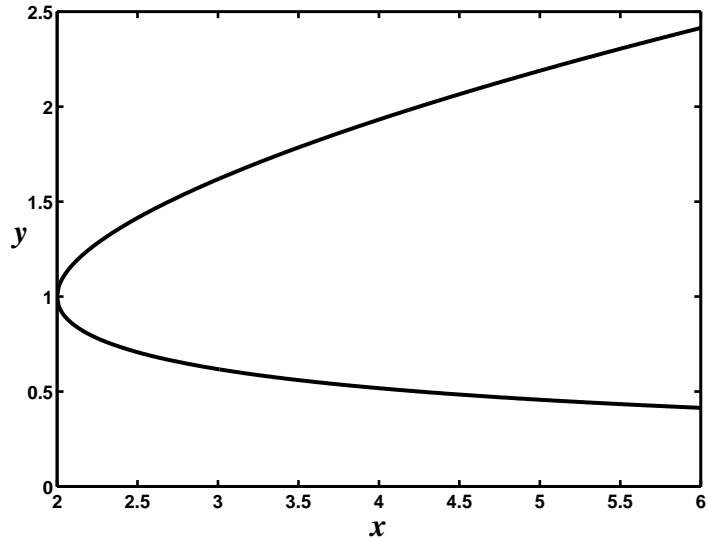}
  \caption{\label{fig1}
  The dispersion relation for quantum oscillations of electron
  gas in plasma based on the results of Ref.~\cite{Dvo}.}
\end{figure}

The dispersion relation which couples the frequency of the
oscillations $\omega$ and the quantum number $k$ reads
\begin{equation}\label{disprelk}
  k^2_{\pm} = \frac{\omega m}{\hbar}
  \left[
    1 \pm
    \left(
      1-4\frac{\omega_p^2}{\omega^2}
    \right)^{1/2}
  \right].
\end{equation}
It is convenient to rewrite Eq.~\eqref{disprelk} in the new
variables $x=\omega/\omega_p$ and $y=k\sqrt{\hbar/2 \omega_p m}$.
Now it is transformed into the form,
\begin{equation}\label{ypm}
  y_{\pm{}}^2=\frac{x}{2}
  \left[
    1 \pm
    \left(
      1-\frac{4}{x^2}
    \right)^{1/2}
  \right].
\end{equation}
We represent the functions $y_{\pm{}}$ versus $x$ on
Fig.~\ref{fig1}.
One can see on Fig.~\ref{fig1} that there are two branches
$y_{\pm{}}$ in the dispersion relation, upper and lower ones. It
is possible to identify the upper branch with a high energy
solution and the lower branch -- with a low energy one.

Generally Eqs.~\eqref{Schrod0} and~\eqref{Uinterac} include the
interaction of the considered electron with all other electrons in
the system. However since we look for the solution in the
pertubative form~\eqref{sol0} and linearize these equations, the
solution to Eqs.~\eqref{pertsol} and~\eqref{disprelk} describe the
interactions of the considered electron only with background
electrons, which correspond to the coordinate independent wave
function $\psi_0$, rather than with electrons which participate in
spherically symmetrical oscillations.

\section{Interaction of two electrons performing oscillations corresponding
to the low energy branch}\label{LE}

In this section we discuss the interaction between two electrons
participating in spherically symmetrical oscillations. This
interaction is supposed to be mediated by a plasmon field
$\varphi$. To account for the electron interaction with a plasmon
one has to include the additional term $e\varphi$ into the
Hamiltonian $H_0$ in Eq.~\eqref{Schrod0}. We also admit that
electron gas oscillations correspond to the lower branch $y_{-{}}$
in the dispersion relation~\eqref{ypm}, i.e. we are interested in
the dynamics of a low energy solution.

A plasmon field distribution can be obtained directly from the
Maxwell equation for the electric displacement field $\mathbf{D}$,
\begin{equation}\label{MaxD}
  (\nabla \mathbf{D})=4\pi\rho,
\end{equation}
where $\rho$ is the free charge density of electrons. As it was
mentioned in Ref.~\cite{Dvo} the vector potential for spherically
symmetrical oscillations can be put to zero $\mathbf{A}=0$.
Therefore only longitudinal plasmons can exist. Presenting the
potential $\varphi$ in the form of the Fourier integral,
\begin{equation}
  \varphi(\mathbf{r},t)=\int
  \frac{\mathrm{d}\omega\mathrm{d}^3\mathbf{p}}{(2\pi)^4}
  \varphi(\mathbf{p},\omega)
  e^{-\mathrm{i}\omega t+\mathrm{i}\mathbf{p}\mathbf{r}},
\end{equation}
we get the formal solution to Eq.~\eqref{MaxD} as
\begin{equation}\label{plasmsol}
  \varphi(\mathbf{r},t)=4\pi\int
  \mathrm{d}t'\mathrm{d}^3\mathbf{r}'
  G(\mathbf{r}-\mathbf{r}',t-t')\rho(\mathbf{r}',t'),
\end{equation}
where
\begin{equation}\label{Greendef}
  G(\mathbf{r},t)=\int
  \frac{\mathrm{d}\omega\mathrm{d}^3\mathbf{p}}{(2\pi)^4}
  \frac{1}{\mathbf{p}^2\varepsilon_l(\mathbf{p},\omega)}
  e^{-\mathrm{i}\omega t+\mathrm{i}\mathbf{p}\mathbf{r}},
\end{equation}
is the Green function for the longitudinal plasmon. In
Eq.~\eqref{Greendef} $\varepsilon_l$ stands for the longitudinal
permittivity. Note that to choose the exact form of the Green
function~\eqref{Greendef} one has to specify passing-by a residue
at $\mathbf{p}=0$. We will regularize the Green function by
putting a cut-off at small momenta. It should be noticed that the
more rigorous derivation of the plasmon Green function is
presented in Ref.~\cite{plasmonprop}.


To relate Eq.~\eqref{plasmsol} with the distribution of electrons
we can take that free charge density is proportional to the
probability distribution, $\rho=e|\psi|^2$. Then, using
Eq.~\eqref{plasmsol} and the Schr\"odinger
equation~\eqref{Schrod0} with the new Hamiltonian $H = H_0 +
e\varphi$ we obtain the closed non-linear equation for the the
electron wave function only.

Now we write down the new Hamiltonian explicitly,
\begin{equation}\label{newHam}
  H=H_0+V,
  \quad
  V=4\pi e^2\int
  \mathrm{d}t'\mathrm{d}^3\mathbf{r}'
  G(\mathbf{r}-\mathbf{r}',t-t')|\psi(\mathbf{r}',t')|^2.
\end{equation}
As we mentioned in Sec.~\ref{MODEL} the term $H_0$ describes the
properties of a single electron including its interactions with
background particles. The second term $V$ in the
Hamiltonian~\eqref{newHam} is responsible for the interactions
between two separated electrons. The scattering of two electrons
is due to this interaction.

It is possible to introduce the additional energy acquired in
scattering of two electrons as
\begin{equation}\label{addE1}
  \Delta E = \langle \psi | V | \psi \rangle =
  4\pi e^2\int
  \mathrm{d}t'\mathrm{d}^3\mathbf{r}'\mathrm{d}t\mathrm{d}^3\mathbf{r}
  \psi^{*{}}(\mathbf{r},t)\psi^{*{}}(\mathbf{r}',t')
  G(\mathbf{r}-\mathbf{r}',t-t')
  \psi(\mathbf{r},t)\psi(\mathbf{r}',t').
\end{equation}
In Eq.~\eqref{addE1} we add one additional integration over $t$
which is convenient for further calculations. To take into account
the exchange effects one has to replace the products of the
electrons wave functions in the integrand in Eq.~\eqref{addE1}
with the symmetric or antisymmetric combinations of the basis wave
functions,
\begin{align}\label{psisymasym}
  \psi(\mathbf{r},t)\psi(\mathbf{r}',t') \to &
  \frac{1}{\sqrt{2}}
  [e^{-\mathrm{i}\omega_1 t-\mathrm{i}\omega_2 t'}
  \psi_{k_1}(\mathbf{r})\psi_{k_2}(\mathbf{r}')\pm
  e^{-\mathrm{i}\omega_2 t-\mathrm{i}\omega_1 t'}
  \psi_{k_2}(\mathbf{r})\psi_{k_1}(\mathbf{r}')],
  \notag
  \\
  \psi^{*{}}(\mathbf{r},t)\psi^{*{}}(\mathbf{r}',t') \to &
  \frac{1}{\sqrt{2}}
  [e^{\mathrm{i}f_1 t+\mathrm{i}f_2 t'}
  \psi_{q_1}^{*{}}(\mathbf{r})\psi_{q_2}^{*{}}(\mathbf{r}')\pm
  e^{\mathrm{i}f_2 t+\mathrm{i}f_1 t'}
  \psi_{q_2}^{*{}}(\mathbf{r})\psi_{q_1}^{*{}}(\mathbf{r}')],
\end{align}
where $k_{1,2}$ and $q_{1,2}$ are the initial and final quantum
numbers, and $\omega_{1,2}$ and $f_{1,2}$ are the initial and
final frequencies of electron gas oscillations. Since the total
wave function of two electrons, which also includes their spins,
should be anti-symmetric, the coordinate wave functions in
Eq.~\eqref{psisymasym} which are symmetrical correspond to
anti-parallel spins. For the same reasons the anti-symmetrical
functions in Eq.~\eqref{psisymasym} describe particles with
parallel spin.

It is worth to mention that in calculating the additional
scattering energy~\eqref{addE1} electrons should be associated
with the perturbative wave function~\eqref{pertsol} rather than
with the total wave function~\eqref{sol0}. However we will keep
the notation $\psi$ for an electron wave function to avoid
cumbersome formulas. Using the basis wave
functions~\eqref{pertsol}, Eqs.~\eqref{addE1}
and~\eqref{psisymasym} as well as the known value of the integral,
\begin{equation}
  \int_0^{\infty}\frac{\mathrm{d}s}{s}\sin(ps)\sin(ks)\sin(qs)
  =
  \frac{\pi}{8}
  [\mathrm{sign}(p+k-q)+\mathrm{sign}(p+q-k)-
  \mathrm{sign}(p+k+q)-\mathrm{sign}(p-k-q)],
\end{equation}
we obtain for the additional energy the following expression:
\begin{align}\label{addE2}
  \Delta E = & \delta(f_1+f_2-\omega_1-\omega_2)
  A_{q_1}^{*{}}A_{q_2}^{*{}}A_{k_1}A_{k_2}
  \frac{e^2\pi^4}{16}
  \int_0^\infty \frac{\mathrm{d}p}{p^2}
  \notag
  \\
  & \times
  \bigg\{
  \frac{1}{\varepsilon_l(p,f_1-\omega_1)}
  [\mathrm{sign}(p+k_1-q_1)+\mathrm{sign}(p+q_1-k_1)
  \notag
  \\
  & -
  \mathrm{sign}(p+k_1+q_1)-\mathrm{sign}(p-k_1-q_1)]\times
  [1 \leftrightarrow 2]
  \notag
  \\
  & \pm
  \frac{1}{\varepsilon_l(p,f_2-\omega_1)}
  [\mathrm{sign}(p+k_2-q_1)+\mathrm{sign}(p+q_1-k_2)
  \notag
  \\
  & -
  \mathrm{sign}(p+k_2+q_1)-\mathrm{sign}(p-k_2-q_1)]\times
  [1 \leftrightarrow 2]
  \bigg\},
\end{align}
where the symbol $[1 \leftrightarrow 2]$ stands for the terms
similar to the terms preceding each of them but with all
quantities with a subscript $1$ replaced with the corresponding
quantities with a subscript $2$.

We suggest that electrons move towards each other before the
collision and in the opposite directions after the collision, i.e.
$k_1=-k_2=k$ and $q_1=-q_2=q$. In this situation the electrons
occupy the lowest energy state (see, e.g.,
Ref.~\cite{LifPit78p191}). Taking into account that the
frequencies and the coefficients $A_k$ are even functions of $k$
[see Eq.~\eqref{disprelk}] we rewrite the expression for the
additional energy in the form,
\begin{align}\label{addE3}
  \Delta E = & \delta(f-\omega)
  |A_{q}|^2 |A_{k}|^2
  \frac{e^2\pi^4}{16}
  \int_0^\infty \frac{\mathrm{d}p}{p^2}
  \frac{1}{\varepsilon_l(p,f-\omega)}
  \notag
  \\
  & \times
  [\mathrm{sign}(p+k-q)+\mathrm{sign}(p+q-k)-
  \mathrm{sign}(p+k+q)-\mathrm{sign}(p-k-q)]^2,
\end{align}
where $\omega=\omega_1=\omega_2$ and $f=f_1=f_2$. It is important
to note that only symmetric wave functions from
Eq.~\eqref{psisymasym} contribute to Eq.~\eqref{addE3}. The
anti-symmetric wave functions convert Eq.~\eqref{addE2} to zero.
It means that only electrons with oppositely directed spins
contribute to the additional energy.

Now we should eliminate the energy conservation delta function.
One can make it in a standard way, $\delta(f-\omega) \to
\text{Time}/(2\pi)$, and then divide the whole expression on the
observation time. Since the energy is concerned in the scattering
one has two possibilities for the momenta, $k+q=0$ or $k-q=0$. We
can choose, e.g., the latter case supposing that $k>0$. In this
situation one has for $\Delta E$ the following expression:
\begin{equation}
  \Delta E = |A_{k}|^4
  \frac{e^2\pi^3}{32}
  \int_0^\infty \frac{\mathrm{d}p}{p^2}
  \frac{1}{\varepsilon_l(p,0)}
  [1-\mathrm{sign}(p-2k)]^2.
\end{equation}
Finally accounting for the step function in the integrand we can
present the additional energy in the form,
\begin{equation}\label{addE4}
  \Delta E = |A_{k}|^4
  \frac{e^2\pi^3}{8}
  \int_0^{2k} \frac{\mathrm{d}p}{p^2}
  \frac{1}{\varepsilon_l(p,0)}.
\end{equation}
Depending on the sign of the the additional energy one has either
effective repulsion or effective attraction.

As we have found in Ref.~\cite{Dvo} the typical frequencies of
electron gas oscillations in our problem $\sim 2\omega_p$, which
is $\sim 10^{13}\thinspace\text{s}^{-1}$. Such frequencies lie
deep inside the microwave region. The permittivity of plasma for
this kind of situation has the following form (see, e.g.,
Ref.~\cite{Gin60}):
\begin{equation}\label{perm}
  \varepsilon_l(\omega)=1-
  \frac{\omega_p^2}{\omega^2+\nu^2},
\end{equation}
where $\nu$ is the transport collision frequency. In
Eq.~\eqref{perm} we neglect the spatial dispersion. It is clear
that in Eq.~\eqref{addE4} one has to study the static limit, i.e.
$\omega \to 0$. Supposing that the density of electrons is
relatively high, i.e. $\omega_p \gg \nu$, one gets that the
permittivity of plasma becomes negative and the interaction of two
electrons results in the effective attraction.

Note that at the absence of spatial dispersion the integral is
divergent on the lower limit. Such a divergence is analogous to
the infrared divergence in quantum electrodynamics. We have to
regularize it by the substitution $0 \to 1/a_D$, where
$a_D=\sqrt{4\pi e^2 n_e/k_B T}$ is the Debye length and $T$ is the
plasma temperature.

Besides the energy $\Delta E$, which is acquired in a collision,
an electron also has its kinetic energy and the energy of the
interaction with background charged particles. For the effective
attraction to take place we should compare the additional negative
energy with the eigenvalue of the operator $H_0$, which includes
both kinetic term and interaction with other electrons and ions.
We suggest that two electrons, which are involved in the
scattering, move towards each other, i.e.
$\omega_1=\omega_2=\omega$. Therefore, if we get that
$2\hbar\omega + \Delta E < 0$, the effective attraction takes
place.

Calculating the integral in Eq.~\eqref{addE4} one obtains the
following inequality to be satisfied for the attraction to happen:
\begin{equation}\label{ineq1}
  x<\frac{\xi^2 \pi^2}{256 \alpha}\frac{n_0 \nu^2 \hbar v}{m \omega_p^4}
  \frac{(y-\tilde{y})}{y^5},
  \quad
  \tilde{y} = \frac{1}{2}\sqrt{\frac{\hbar\omega_p}{2 m v^2}},
\end{equation}
where $v = \sqrt{k_B T/m}$ is the thermal velocity of background
electrons in plasma and $k_B \approx 1.38 \times
10^{-16}\thinspace\text{erg/K}$ is the Boltzmann constant. In
Eq.~\eqref{ineq1} we use the variables $x$ and $y$ which were
introduced in Sec.~\ref{MODEL}. We also assume that $A_k^2 =
n_c/k^2$, where $n_c$ is the density in the central region [see
Eq.~\eqref{pertsol}]. As it was predicted in our work~\cite{Dvo},
$n_c$ should have bigger values that the density on the rim of the
system. We can also introduce two dimensionless parameters, $\xi$
and $\alpha$, to relate the density in the center, the density of
free electrons, and the rim density: $n_c = \xi n_0$ and $n_e =
\alpha n_0$.

We suppose that $\alpha \sim 0.1$. It is known (see, e.g.,
Ref.~\cite{PEv3}) that the number density of free electrons in a
spark discharge is equal to
$10^{16}-10^{18}\thinspace\text{cm}^{-3}$. Approximately the same
electrons number densities are acquired in a streak lightning. If
one takes the rim density equal to the Loschmidt constant $n_0=2.7
\times 10^{19}\thinspace\text{cm}^{-3}$, we arrive to the maximal
value of $\alpha \sim 0.1$. In this case mainly the collisions
with neutral atoms would contribute to the transport collision
frequency. One can approximate this quantity as $\nu \approx v
\sigma n_c$, where $\sigma = \pi R^2$ is the characteristic
collision cross section and $R \sim 10^{-8}\thinspace\text{cm}$ is
the typical atomic radius. Note that the process in question
happens in the center of the system. That is why we take the
central number density $n_c$ in the expression for $\nu$.

Finally we receive the following inequality to be satisfied for
the effective attraction between two electrons to take place:
\begin{gather}
  x < K
  \frac{\tau^{3/2}\xi^4}{\alpha^3}\frac{1}{y^5}
  \left(
    y-y_0
    \frac{\alpha^{1/4}}{\tau^{1/2}}
  \right),
  \quad
  y_0 = \frac{1}{2}\sqrt{\frac{\hbar\omega_0}{2 m v_0^2}}
  \approx 0.63,
  \notag
  \\
  \label{ineq2}
  K = \frac{\pi^2}{256}\frac{\hbar n_0^3 v_0^3\sigma^2}{m \omega_0^4}
  \approx 1.1 \times 10^{-11},
\end{gather}
where $\tau = T/T_0$. In Eq.~\eqref{ineq2} we take the reference
temperature $T_0=700\thinspace\text{K}$. Now one calculates other
reference quantities like $v_0$ and $\omega_0$ as $v_0 = \sqrt{k_B
T_0/m} \approx 10^7\thinspace\text{cm/s}$ and $\omega_0 = \sqrt{4
\pi e^2 n_0/m} \approx 2.9 \times 10^{14}\thinspace\text{s}^{-1}$.

As one can see on Fig.~\ref{fig1} [see also Eq.~\eqref{ypm}] the
parameter $y$ is always more than $1$ for the upper branch. It
means that the inequality in Eq.~\eqref{ineq2} can never be
satisfied. On the contrary, there is a possibility to implement
the condition~\eqref{ineq2} for the lower branch on
Fig.~\ref{fig1}. It signifies the existence of the effective
attraction between electrons.

\begin{figure}
  \centering
  \includegraphics[scale=.4,angle=270]{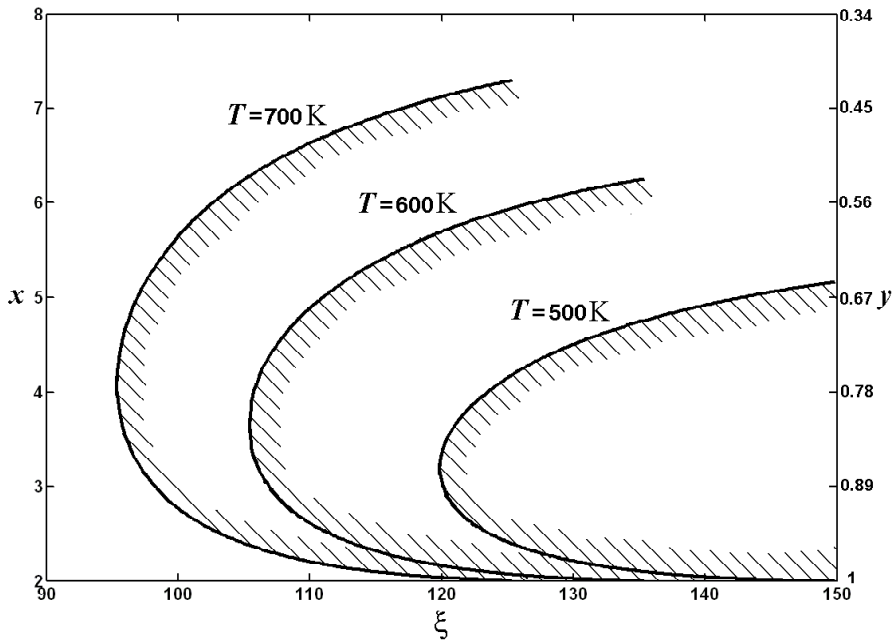}
  \caption{\label{fig2}
  The allowed regions in the $(\xi,x)$ plane
  for the existence of the effective attraction
  between two electrons performing quantum oscillations corresponding
  to the lower branch on Fig.~\ref{fig1}. We also add one more
  vertical axis on the right hand side demonstrating the behaviour
  of the quantum number $k$. Three zones for the temperatures
  $T=500\thinspace\text{K}$, $600\thinspace\text{K}$ and
  $700\thinspace\text{K}$
  are shown here.}
\end{figure}

On Fig.~\ref{fig2} we plot the allowed regions where the effective
attraction happens.
This plot was created for the constant value of $\alpha = 0.1$. We
analyze the condition~\eqref{ineq2} numerically. The calculation
is terminated when we reach the point $x=2$. It has the physical
sense since oscillations cannot exist for $x<2$ according to
Eq.~\eqref{ypm} (see also Fig.~\ref{fig1}).

One can notice that always there is a critical value of the
central density. It is equal to $120 n_0$, $105 n_0$ and $95 n_0$
for the temperatures $T=500\thinspace\text{K}$,
$600\thinspace\text{K}$ and $700\thinspace\text{K}$ respectively.
It means that in the center of the system there is an additional
pressure of about $100\thinspace\text{atm}$.

One has to verify the validity of the used approach checking that
$y>\tilde{y}$ in Eq.~\eqref{ineq1}. It would signify that that the
the quantum number $k$ is more than $1/2a_D$. The values of
$\tilde{y}$ are $0.42$, $0.38$ and $0.35$ for
$T=500\thinspace\text{K}$, $600\thinspace\text{K}$ and
$700\thinspace\text{K}$ respectively. Comparing these values with
the curves on Fig.~\ref{fig2} one can see that the condition
$y>\tilde{y}$ is always satisfied for all the allowed zone of the
effective attraction existence.

For the effective attraction to happen one should have the
negative values of $\varepsilon_l(0)$ in Eq.~\eqref{addE4}. Using
Eq.~\eqref{perm} one can conclude that the following inequality
should be satisfied: $\nu < \omega_p$ or equivalently $\xi < 3.3
\times 10^{3} \sqrt{\alpha/\tau}$. For $\alpha = 0.1$ and
temperatures $T=500\thinspace\text{K}$, $600\thinspace\text{K}$
and $700\thinspace\text{K}$ we get the critical values of $\xi$ as
$1.2 \times 10^3$, $1.1 \times 10^3$ and $1.0 \times 10^3$
respectively. One sees on Fig.~\ref{fig2} that all the curves lie
far below the critical values of $\xi$. Therefore the condition
$\varepsilon_l < 0$ is always satisfied.

\section{Summary and discussions}\label{CONCL}

In conclusion we mention that we studied spherically symmetrical
quantum oscillations of electron gas in plasma which correspond to
a low energy solution to the Schr\"{o}dinger equation~\cite{Dvo}.
We suggested that two independent plasma excitations,
corresponding to two separate electrons, can interact via exchange
of a plasmon. Since the typical oscillations frequencies are very
high and lie in the deep microwave region the plasma permittivity
is negative and the interaction of two electrons result in the
appearance of an attraction. For the this attraction to be actual
we compared it with the kinetic energy of electrons and the energy
of interaction of electrons with other background charged
particles. The total energy of an electrons pair turned out to be
negative for the lower branch in the dispersion relation (see
Fig.~\ref{fig1} and Ref.~\cite{Dvo}). Although we did not
demonstrate it explicitly, using the same technique as in
Sec.~\ref{LE} one can check that the total energy of electrons is
always positive for the upper branch. It means that even
additional negative energy does not cause the attraction of
electrons. It is interesting to mention that to obtain the master
equations~\eqref{Schrod0} and~\eqref{Uinterac} in Ref.~\cite{Dvo}
we neglected the exchange effects between an electron which
corresponds to the wave function $\psi$ and the rest of background
electrons (see also Ref.~\cite{KuzMak99}). This crude
approximation is valid if the superconductivity takes place
because in this case the friction, i.e. the interaction, between
oscillating and background electrons is negligible.

The proposed plasmon superconductivity happening at spherically
symmetrical oscillations of electrons could be implemented in a
low energy ball lightning. Theoretical and experimental studies of
ball lightnings have many years history (see, e.g.,
Ref.~\cite{Tur98}). This natural phenomenon, happening mainly
during a thunderstorm, is very rare and we do not have so many its
reliable witnesses. There were numerous attempts to generate in a
laboratory stable structures similar to a ball lightning. Many
theoretical models aiming to describe the observational data were
put forward. In putting forward a ball lightning model one should
try to explain these properties on the basis of the existing
physical ideas without involvement of extraordinary concepts (see,
e.g. Ref.~\cite{Rab99}) though they look very exciting. However
none of the available models could explain all the specific
properties of a fireball.

Among the existing ball lightning models one can mention the
aerogel model~\cite{Smi93}. According to this model fractal fibers
of the aerogel can form a knot representing the skeleton of a ball
lightning. Using this model it is possible to explain some of the
ball lightning features. The interesting model of a ball lightning
having complex onion-like structure with multiple different layers
was put forward in Ref.~\cite{Tur94}. However in this model an
external source of electrical energy is necessary for a ball
lightning to exist for a long time. The hypothesis that a ball
lightning is composed of molecular clusters was described in
Ref.~\cite{Sta85}. This model can explain the existence of a low
energy ball lightning. In Ref.~\cite{RanSolTru00} the
sophisticated ball lightning model was proposed, which is based on
the non-trivial magnetic field configuration in the form of closed
magnetic lines forming a knot. In frames of this model one can
account for the relatively long life time of a ball lightning. It
is worth to be noticed that though the properties of low energy
fireballs can be accounted for within mentioned above models they
are unlikely to explain its regular geometrical shape. The author
of the present work has never observed a ball lightning, however
according to the witnesses a fireball resembles a regular sphere.
There are many other ball lightning models which are outlined in
Ref.~\cite{Smi88}.

The most interesting ball lightning properties are (see, e.g.,
Ref.~\cite{Bar80})
\begin{itemize}
  \item There are both high and low energy ball lightnings.
  The energy of a low energy ball lightning can be below several
  hundred~kJ. However the energy of a high energy one can be
  up to $1\thinspace\text{MJ}$.
  \item There are witnesses that a ball lightning can penetrate
  through a window glass. Sometimes it uses existing microscopic
  holes without destruction of the form of a ball lightning. It means
  that the actual size of a ball lightning is rather small and its
  visible size of several centimeters is caused by some secondary effects.
  Quite often a ball lightning can burn tiny holes inside materials
  like glass to pass through them. It signifies that the internal
  temperature and pressure are very high in the central region.
  \item A fireball is able to follow the electric field lines.
  It is the indirect indication that a ball lightning has
  electromagnetic nature, e.g., consists of plasma.
  \item A ball lightning has very long life-time, about several
  minutes. In Sec.~\ref{INTR} we mentioned that unstructured
  plasma without external energy source looses its energy and
  recombines extremely fast. If we rely on the plasma models of a
  ball lightning, it means that either there is an energy source
  inside of the system or plasma is structured and there exists a
  mechanism which prevents the friction in the electrons motion.
  \item In some cases there were reports that a ball lightning
  could produce rather strong electromagnetic radiation even in
  the X-ray range. It might signify that energetic processes, e.g.
  nuclear fusion reactions, can happen inside a fireball.
\end{itemize}

Quantum oscillations of electron gas in plasma~\cite{Dvo} can be a
suitable model for a ball lightning. For example, it explains the
existence of two types of ball lightnings, low and high energy
ones. Our model accounts for the very small $\sim
10^{-7}\thinspace\text{cm}$ and dense central region as well as
indirectly points out on the possible microdose nuclear fusion
reactions inside of a high energy ball lightning.

In the present paper we suggested that superconductivity could
support long life-time of a low energy fireball and described that
this phenomenon could exist in frames of our model. It is unlikely
that nuclear fusion reactions can take place inside a low energy
ball lightning, i.e. this type of a ball lightning cannot have an
internal energy source. Moreover as we mentioned in
Ref.~\cite{Dvo}, electrons participating in spherically
symmetrical oscillations cannot emit electromagnetic radiation. It
means that the system does not loose the energy in the form of
radiation. As a direct consequence of our model we get the regular
spherical form of a fireball gratis. In Sec.~\ref{LE} we made the
numerical simulations for the temperature range
$500-700\thinspace\text{K}$. According to the fireball
witnesses~\cite{Bar80} some low energy ball lightnings do not
combust materials like paper, wood etc. The combustion temperature
of these materials lie in the mentioned above range. It justifies
our assumption.

Note the a ball lightning seems to be a many-sided phenomenon and
we do not claim that our model explains all the existing
electrical atmospheric phenomena which look like a ball lightning.
However in our opinion a certain class of fireballs with the above
listed properties is satisfactory accounted for within the model
based on quantum oscillations of electron gas in plasma.

\begin{acknowledgments}
The work has been supported by the Conicyt (Chile), Programa
Bicentenario PSD-91-2006. The author is very thankful to Sergey
Dvornikov  and Viatcheslav Ivanov for helpful discussions.
\end{acknowledgments}

\end{document}